\documentclass[preprint,aps]{revtex4} 
\usepackage{graphicx}
\usepackage{dcolumn}
\usepackage{bm}%
\usepackage{natbib}

\begin{document}

\title{\bf Magnetic field inhibits the conversion of neutron stars to quark stars} 

\author{Ritam Mallick$^{a,b}$\footnote{Email : ritam@bosemain.boseinst.ac.in}, 
Sanjay K. Ghosh$^{a,b}$\footnote{Email : sanjay@bosemain.boseinst.ac.in} and 
Sibaji Raha$^{a,b}$\footnote{Email : sibaji@bosemain.boseinst.ac.in; Corresponding author}}

\affiliation{$^a$Department of Physics; Bose Institute;
 93/1, A.P.C Road; Kolkata - 700009; INDIA}
\affiliation{$^b$Centre for Astroparticle Physics and
Space Science; Bose Institute; Block - EN, Sector V; Salt Lake; Kolkata - 700091;
INDIA}

\begin{abstract}
Neutron stars provide a natural laboratory to test some unique implications
of Quantum Chromodynamics (QCD)- the underlying theory of strong interactions- at
extreme conditions of very high baryon density. It has been suggested that the 
true ground state of QCD is strange quark matter, 
and, consequently, neutron stars should convert to strange quark stars under
suitable conditions. Substantial efforts have been, and are being, spent in 
studying the details of such conversion. In this report, we show that the
presence of high magnetic field, an essential feature of neutron stars, 
strongly inhibits the conversion of neutron stars to bare quark stars.
\end{abstract}

\maketitle

Neutron stars (NS) are compact objects formed as an aftermath of supernovae, 
resulting from the gravitational collapse of massive stars which have exhausted 
their nuclear fuel. The central density of NS is believed to be very
high ($6-10$ times nuclear saturation density, depending on the total mass).
They are also characterized by very high magnetic fields ($10^{10}-10^{12}$ 
G) at the surface. While the origin of such magnetic fields in NS 
remains unclear (trapping of all magnetic lines of force during the 
collapse \cite{key-1} can barely yield $10^8-10^9$ G), the observed 
slow-down rate of pulsars, which are rotating NS, do require such 
magnetic fields. Thus, the canonical picture of the classical pulsar mechanism 
involves \cite{key-1} a magnetic dipole at the centre of a rotating NS. 
The maximum mass of NS is somewhat model dependent, but in any case, of the 
order of the Chandrasekhar limit.

In the classical picture, {\it {i.e.}}, before the advent of QCD, it was 
believed that the only compact astrophysical objects were the white dwarfs
and NS; anything other than  these two classes would end up
in black holes. However, QCD predicts the existence of a new class of compact
stars, called quark stars \cite{key-2} (QS), which again obey a 
maximum mass limit \cite{key-3} of the order of Chandrasekhar limit. 
It was proposed \cite{key-4,key-5} some time ago that the stable strange 
quark matter (SQM) was the {\it true} ground state of QCD and as such, all NS ought 
to evolve, under suitable conditions, to this more stable configuration of QS, 
more correctly, strange stars (SS). Some tentative SS candidates are 
the compact 
objects associated with X-ray bursters GRO $J1744-28$ \cite{key-6}, 
SAX $J1808.4-3658$ \cite{key-7} and X-ray pulsars Her $X-1$ \cite{key-8}.
It has also been argued in the literature that sub-millisecond pulsars, if
they actually exist, must be QS; the NS would not be stable at such high
rotation speeds.

There could of course be primodial SS, which could provide a natural 
explanation of the cosmological dark matter (CDM) within the standard 
model \cite{key-9}. For the present purpose, we do not address them 
but rather concentrate on the evolution of NS to SS. There are 
several plausible 
scenarios where NS could convert to quark stars (ultimately SS), 
through a "seed" of external SQM \cite{key-10}, or
triggered by the rise in the central density due to sudden 
spin-down \cite{key-11}. It has been argued in the 
literature that such a conversion would not only provide a natural engine for 
the gamma ray bursts (GRB) but can also explain their observed beaming 
property {\cite{key-12}.

Starting from the original work of MIT group \cite{key-10,key-13}, a 
number of authors has studied the details of the conversion of NS to QS/SS; 
they have been summarized in our previous chapters and works
\cite{key-14,key-15}. 
(For the sake of brevity, it is not possible to cite all of 
them here.) Suffice is to say, most of them have ignored some essential 
features of NS, namely the rotational effect, the general relativistic 
(GR) effect 
and, most importantly, the topic of this chapter, the role of the strong 
magnetic field. We have systematically tried to incorporate these effects in a 
series of works. In Ref.~14, we have shown that the 
conversion process is most 
likely a two-step process. The first step involves the conversion of NS to a 
two-flavour QS, occurring on millisecond timescales. The second step comprises 
the conversion from two-flavour quark matter to the stable three-flavour SQM, 
through weak interaction, the corresponding time scale being of the 
order of $100$ 
seconds. In our more recent work \cite{key-15}, we have exhibited the major 
role played by the GR effects in the rotating stars; the conversion 
fronts propagate 
with different velocities along different radial directions, 
a finding which could 
not have been anticipated from newtonian or  special 
relativistic (SR) analyses. 
We saw that the velocity of the conversion front increased outward and engulfed 
all the matter, converting the NS to a bare QS.

In the present work, we report the startling finding that incorporation
of the magnetic field drastically alters the scenario. The formalism of the
calculation, as reported in Ref.~15, remains unchanged and we need not discuss 
them in detail here; only the bare essentials are mentioned below. 

We start with the metric \cite{key-16}
\begin{eqnarray}
ds^2 = -e^{\gamma+\rho}dt^2 + e^{2\alpha}(dr^2+r^2d\theta^2) + 
e^{\gamma-\rho}r^2 sin^2\theta(d\phi-\omega dt)^2
\end{eqnarray}
describing the structure of the star, with the four gravitational potentials 
$\alpha, \gamma, \rho$ and $\omega$, which are functions of $\theta$ and $r$ 
only. The Einstein's equations for the potentials are solved through the 
{\bf 'rns'} code \cite{key-17}, with the input of an equation of state 
(EOS) and a central density. The nuclear matter phase is described by 
the non-linear Walecka model \cite{key-18} and the quark phase 
by the MIT 
Bag model \cite{key-19}. We, as usual, treat both the nuclear 
and quark phases 
as ideal fluids. The operative equations, then, are the continuity and Euler's
equations; (for details, please see Ref. 15) 
\begin{eqnarray}
\frac{1}{\varpi}\left(\frac{\partial\epsilon}{\partial\tau}
+v\frac{\partial\epsilon}{\partial r}\right)+
\frac{1}{W^{2}}\left(\frac{\partial v}{\partial r}
+v\frac{\partial v}{\partial\tau}\right)+\frac{2v}{r}+\frac{v}{r}cot\theta  
= -v \left(\frac{\partial\gamma}
{\partial r} + \frac{\partial \alpha}{\partial r}\right)\\
\frac{1}{\varpi}\left(\frac{\partial p}{\partial r}+
v\frac{\partial p}{\partial\tau}\right)+
\frac{1}{W^{2}}\left(\frac{\partial v}{\partial\tau}+
v\frac{\partial v}{\partial r}\right) 
=\frac{1}{2}\left(A\frac{\partial\gamma}
{\partial r} + B\frac{\partial \rho}{\partial r}+C\frac{\partial \omega}
{\partial r}+E\right),
\end{eqnarray}
where, $v$ is the r.m.s. velocity, $\varpi$ is the enthalpy and $W$
is the inverse of the Lorentz factor.

The presence of magnetic field alters these equations in a straightforward
manner. The continuity equation is unaffected but the Euler's equation picks up an
additional term, corresponding to the  magnetic pressure 
$\frac{B^2}{8\pi}$ \cite{key-20}. In the above Euler's equation, we
replace pressure $p$ by $p^/ = p + \frac{B^2}{8\pi}$.
With this modification, the calculation proceeds exactly as in Ref.~15.

In keeping with the canonical picture, we assume that the magnetic field 
of the NS is due to a dipole at the centre of the star. At any point $(r, \theta)$,
the field due to the dipole in the radial direction is given by
\begin{equation}
B=\frac{\mu_0 m}{2\pi r^3} cos\theta,
\end{equation}
where $\mu_0$ is the permeability in free space and $m$ is 
the magnetic moment. The dipole is assumed to be along the polar direction 
and $\theta$ is the angle with the vertical axis. We focus mainly on the range
$B_{surface}\sim10^8-10^{12} G$.

We now proceed exactly as in \cite{key-15}. We define $v$ as the front 
velocity in the nuclear matter rest frame and $n=\frac{\partial p}
{\partial\epsilon}$ is the square of the effective sound speed in the medium. 
$d\tau$ and $dr$ are connected by the relation
\begin{equation}
\frac{d r}{d \tau}=vG
\end{equation}
where $G$ is given by
\begin{equation}
G=\sqrt{\frac{e^{\gamma+\rho}-e^{\gamma-\rho}r^2\omega^2 sin^2\theta}{e^{2\alpha}}}
\end{equation}
The other parameters in Eq. 2 and 3 are defined as 
\begin{eqnarray}
A=\frac{v\omega r sin\theta}{C1}-1; \hspace {0.1 in} E=\frac{2\omega^2 r sin\theta}{C1}+\frac{2\omega^2 e^{\gamma-\rho} r sin\theta}{A1}; \nonumber\\
B=\frac{B1}{A1}-\frac{v\omega r sin\theta}{C1}; \hspace {0.1 in} C=\frac{2\omega e^{\gamma-\rho} r^2 sin\theta}{A1}+\frac{vrsin\theta}{C1} \nonumber 
\end{eqnarray}
where 
\begin{eqnarray}
A1=e^{\gamma+\rho}-e^{\gamma-\rho}r^2\omega^2 sin^2\theta; \nonumber \\ 
B1=-e^{\gamma+\rho}-e^{\gamma-\rho}r^2\omega^2 sin^2\theta;  \nonumber \\
C1=\sqrt{r^2\omega^2sin^2\theta-e^{2\rho}}; \nonumber 
\end{eqnarray}

After a bit of algebra, very similar to that in Ref.~15, we can write the 
generic differential equation for the velocity $v$ of the conversion front 
\begin{equation}
\frac{\partial v}{\partial r} = \frac{W^2v[K+K1-K2]}{2[v^2(1+G)^2-n(1+v^2G)^2]}.
\end{equation}
where,
\begin{eqnarray}
K=2n(1+v^2G)\left({\frac{\partial \gamma}{\partial r} + 
\frac{\partial \alpha}{\partial r}+\frac{2}{r} +\frac{cot\theta}{r}
}\right) \nonumber \\
K1=(1+G)\left(A\frac{\partial \gamma}{\partial r} + B\frac{\partial \rho}{\partial r}
+C\frac{\partial \omega}{\partial r}+E \right) \nonumber \\
K2=\frac{(1+G)(1+v^2G)}{4\pi^2\varpi}.\frac{{\mu_0}^2m^2cos^2\theta}{r^7}
\end{eqnarray}
Integrating eqn. (5) over $r$ from the center to the surface, we obtain the 
propagation velocity of the front as a function of $r$. 

We perform our calculation for a NS with a central density of $7$ times the 
nuclear saturation density, (the corresponding Keplerian rotation rate being 
$0.89 \times 10^{4} sec^{-1}$). For the sake of comparison, we also study a slowly
rotating star, with rotational velocity of $0.4 \times 10^{4} sec^{-1}$.

\begin{figure}
\vskip 0.2in
\centering
\includegraphics[width=4.0in]{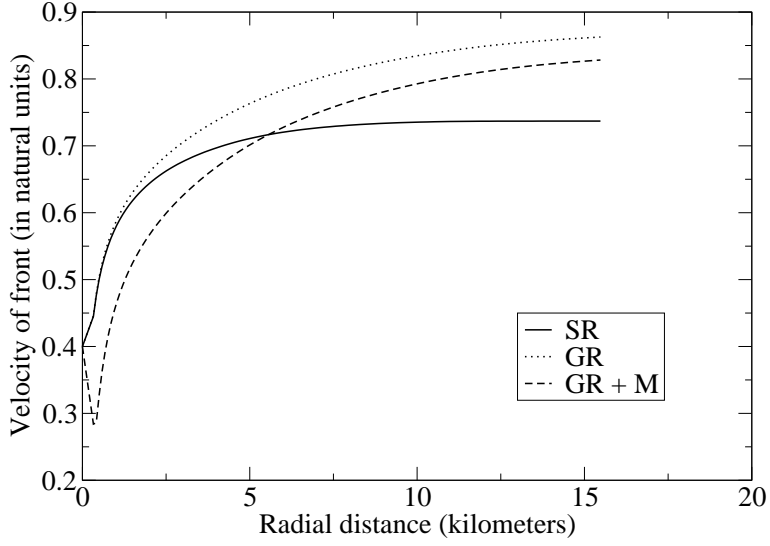}
\caption{Variation of velocity of the conversion front along the equatorial
direction of the star for three different cases, namely SR, GR and GR+M.}
\end{figure}

In fig. 1, we show the propagation velocity of the conversion front 
along the equatorial direction for the Keplerian star. The solid curve denotes 
the SR calculation, the dotted curve the GR result (both reported 
earlier \cite{key-15}) and the dash-dotted curve the result of the present 
calculation, including the GR and magnetic field effects (GR+M). The surface 
magnetic field is $\sim 5 \times 10^{10} G$. Fig. 2 shows the relative comparison 
between the slow star and the Keplerian star. Self consistency condition
yields that the surface magnetic field for the slow star is slightly 
higher $\sim 5.5 \times 10^{10} G$.

\begin{figure}
\vskip 0.2in
\centering
\includegraphics[width=4.0in]{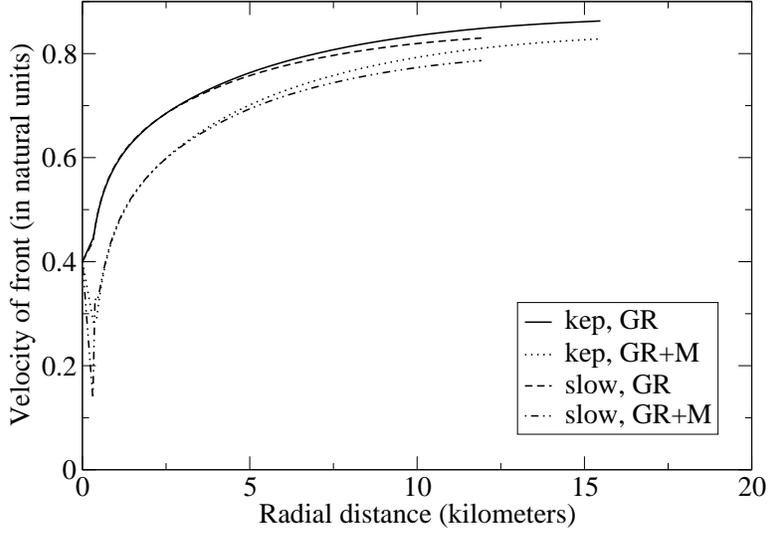}
\caption{Variation of velocity of the conversion front along the equatorial
direction of the star rotating with two different velocities for two cases,
GR and GR+M.}
\end{figure}

\begin{figure}
\vskip 0.2in
\centering
\includegraphics[width=4.0in]{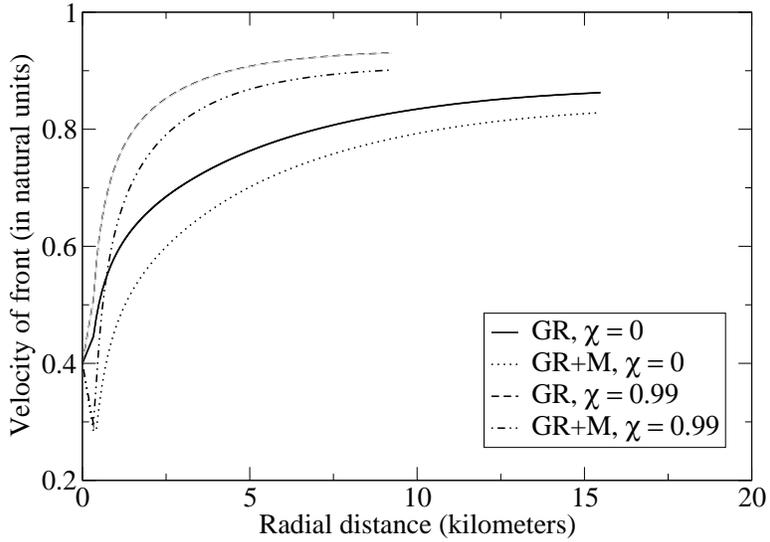}
\caption{Variation of velocity of the conversion front along the radial
direction (polar and equatorial) of the star rotating with Keplerian velocity.}
\end{figure}

From figs. 1 and 2, we clearly notice that near the centre of the star, the
magnetic field produces a 'braking' effect. Only at large radial distances 
does the GR effect start to dominate and accelerate the conversion front.
The effect of the rotational speed is quantitative; we find qualitatively 
similar behaviours both for slowly rotating and Keplerian stars. Fig. 3 shows the 
velocity of the conversion front along the polar and equatorial directions
for the Keplerian star. $\chi \equiv cos\theta$, defined {\it w.r.t} 
the vertical axis of the star.

This 'braking' effect  is naturally due to the additional pressure 
of the magnetic field. Our detailed calculation shows that for surface
field less than $10^{10} G$, the 'braking' effect on the conversion front
is not at all discernible; the GR effect dominates from the very beginning and the 
entire NS may get converted to a bare QS/SS. For higher magnetic fields, 
well within the canonical range, the effect however is much more drastic, 
as shown in fig. 4. 

\begin{figure}
\vskip 0.2in
\centering
\includegraphics[width=4.0in]{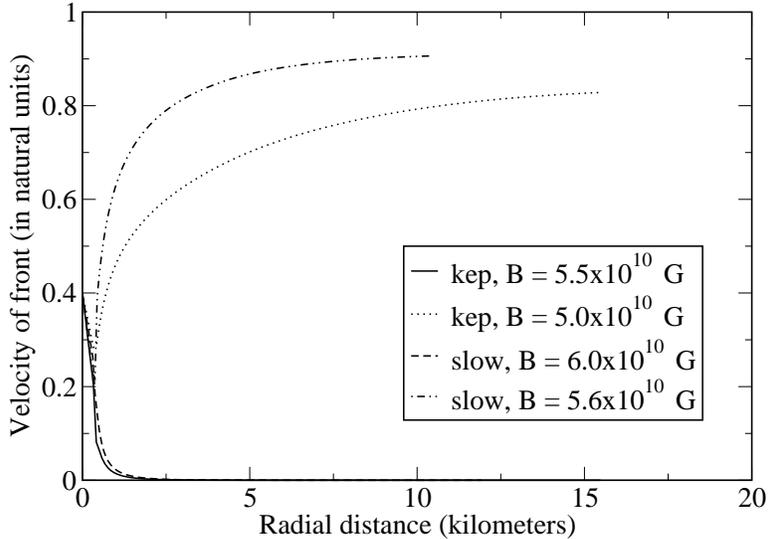}
\caption{Variation of the velocity of the conversion front showing the 'braking'
effect due to the magnetic field for different cases.}
\end{figure}

Fig. 4 shows the results for both slowly rotating as well as Keplerian stars, 
along the equatorial radii. We see that for a surface magnetic 
field of $5.5 \times 10^{10} G$ along the equatorial direction, 
the velocity drops sharply and becomes almost zero at a distance of $3 km$ 
(where equatorial radius is $15 km$) from the centre for the Keplerian star. 
Comparable and qualitatively similar conclusions are also obtained for the 
slow star. In the polar direction, magnetic filed necessary to stall the front 
is slightly smaller than that in the equatorial direction. If the magnetic 
field 
is of the order of $10^{12} G$ at the surface, the quark core shrinks within a 
distance of $\sim 100 m$. For still larger fields, the front fails to start 
propagating. 

The implications of these results are indeed far-reaching. We are forced to 
conclude that the conversion to QS/SS is strongly inhibited in classical NS
characterized by large magnetic fields. In particular, it is obvious that 
while these NS can at best convert to hybrid stars, with a small quark core, 
the magnetars (NS with very high magnetic fields) have no chance to convert to 
QS/SS. Only very young NS, carrying small magnetic fields, have a chance to 
convert to QS/SS.

For completeness, we would like to mention that the conclusion of this 
study does not depend much on the assumption of a dipole magnetic field.
We have also employed an alternate field configuration which has been used 
by some authors \cite{key-21} in recent times. The conclusion remains 
unaltered.

\acknowledgments{R.M. thanks Council of Scientific \& Industrial Research, New Delhi for financial support. 
S.K.G. and S.R. thank Department of Science \& Technology, Govt. of India for support under the IRHPA scheme.}

\end{document}